\documentclass[10pt]{iopart}

\usepackage{mathrsfs}
\usepackage{iopams}
\usepackage{cite}
\usepackage{epsfig}
\usepackage{graphicx}
\usepackage{color}

\def\3nab{\tilde{\nabla}}

\def\be {\begin{equation}}
\def\ee {\end{equation}}
\def\ba {\begin{eqnarray}}
\def\ea {\end{eqnarray}}

\newcommand{\bra}[1]{\left(#1\right)}
\newcommand{\bras}[1]{\left[#1\right]}

\newcommand{\sfr}[2]{{\textstyle\frac{#1}{#2}}}

\newcommand{\E}{{\mathcal E}}

\newcommand{\barray}{\begin{array}}
\newcommand{\earray}{\end{array}}
\newcommand{\N}{N}

\newcommand{\del}{\nabla}

 \newcommand{\nab}{\nabla}

\newcommand \ep {\epsilon}
\newcommand{\LB}{\left(}
\newcommand{\RB}{\right)}

\newcommand{\al}{\alpha_0}
\newcommand{\bl}{\beta_0}
\newcommand{\Tf}{\mathcal{T}}
\newcommand{\Td}{{\Upsilon}}

\newcommand{\mb}{\overline{m}}
\newcommand{\ov}{\overline}
\newcommand{\dum}{Tweedledum }
\newcommand{\dee}{Tweedledee }
\newcommand{\udot}{{\mathcal A}}

\begin{document}

 \title{ Transferring Energy in General Relativity}
\author{Rituparno Goswami\dag, George F. R. Ellis\ddag}
\address{\dag Astrophysics and Cosmology Research Unit, School of Mathematics, Statistics and Computer Science, University of KwaZulu-Natal, Private Bag X54001, Durban 4000, South Africa.}
\address{\ddag Department of Mathematics and Applied Mathematics and ACGC, University of Cape Town,
Cape Town, Western Cape, South Africa.}
 \date{\today}
 \eads{goswami@ukzn.ac.za, George.Ellis@uct.ac.za}

\begin{abstract}
A problem in general relativity is, how the gravitational field can transfer energy and momentum between different distant places. The issue is that matter stress tensor is locally conserved, with no explicit interaction with the free gravitational field, which is represented by the Weyl tensor. In this paper we explicitly construct an interaction tensor for free gravity and matter, that depicts the interplay between the energy momentum tensor of free gravity, which is taken to be the symmetric two index square root of Bell-Robinson tensor, and matter. This is examined both in the case of Coulomb-like Petrov type D spacetimes and radiation like Petrov type N spacetimes, where a unique square root exists. The first case generalises the  \dum and \dee  thought experiment regarding gravitational induction in Newtonian gravity to general relativistic scenarios, and the second gives a proposal for how gravitational radiation can transfer energy and momentum between distant objects separated by a vacuum. 
\end{abstract}


\section{Introduction}

Since the birth of general relativity, the problem which perpetually eluded a satisfactory solution was the exact mechanism through which energy can be transferred from one body to another via gravitational interactions. In electromagnetism this problem is well understood as the electromagnetic energy can be transferred across empty space via induction or radiation, enabled by the electromagnetic field $F_{ab}$. One would then naturally expect the same in case of gravity.\\

In Newtonian gravity, where there cannot be any gravitational waves, the effects of gravitational induction was well understood both theoretically and observationally (for example the tidal effect due to the gravitational induction in the earth moon system). Bondi presented a beautiful thought experiment with two gravitationally interacting bodies and demonstrated, how the internal energy of one body can be transferred to the other via gravitational interaction through the intervening vacuum \cite{bondi,jvn} (see Section 2).\\

In general relativity, the same issue arises both as regards gravitational induction (where the magnetic part of the Weyl tensor vanishes) and gravitational radiation (where both the electric and magnetic parts of the Weyl tensor are non-zero). 
The point here is that the energy-momentum tensor for the matter present is conserved, without any reference to the Weyl tensor $C_{abcd}$, which acts as a free gravitational field conveying energy and momentum across the intervening vacuum. 
One approach has been via stress-energy-momentum pseudotensors, for example the Landau-Lifshitz  pseudotensor \cite{LL}; but as they are not tensors, they can be made to vanish by a suitable choice of coordinates. Detailed investigations concerning the interaction between matter and the gravitational field in the linear case were performed by \cite{hob1,hob2,hob3}, to arrive at new energy-momentum and spin-angular-momentum pseudo-tensors that have very sensible physical properties in the harmonic gauge and are gauge-invariant for gravitational waves. However issues of uniqueness arise \cite{hob4}. A second approach in the case of gravitational radiation is that of Isaacson  \cite{IsaacsonI}, \cite{IsaacsonII}, based on averaging small wavelength oscillations associated with a  gravitational wave to give a backreaction effect on the background spacetime. This is a special case of the more general issue of averaging and backreaction in spacetime \cite{Ell84}. It is an interesting and useful approach, but is again coordinate dependent. \\

What would be preferable if possible, is to use a covariant approach that recognizes the role of the Weyl tensor as the free gravitational field conveying gravitational effects from one place to another \cite{Ell71,EllMaMac}, as proposed by Trautmann, Pirani, Sachs, Trumper, Penrose, Newman, and others, in analogy with the way the electric field does so in the case of electromagnetism \cite{MaaBass}. The way this works out in various cases, including a slow-motion binary system made of nonspinning point particles leading to outgoing quadrupole gravitational waves, is explored by Nicholls et al \cite{Nicholls} and Zhang et al \cite{Zhang}.  {The question we wish to address is ``How does free gravity, as just characterised, interact with matter?'' We wish to emphasise here that this question is NOT the same as ``How does gravity interact with matter?'' 
	The answer to that question is given by the Einstein equations $G_{ab} = T_{ab}$ (matter curves spacetime) and the conservation equations $T^{ab}_{\,\,\,\,\,\,;b}=0$ (spacetime tells matter how to move, via the covariant derivative in that equation). Neither of these equations explicitly involves the Weyl tensor. As in the case of the electromagnetic field, the interaction we are interested in is of a non-local nature, but is mediated by local equations.}\\

The biggest hurdle to such an approach was the absence of any satisfactory energy momentum tensor for free gravity (gravity in the absence of matter). Roger Penrose suggested that it should be constructed entirely via the Weyl tensor, which describes the free gravity on the manifold, but did not give a specific formula. There were several earlier unsuccessful attempts and recently it was proposed by Clifton, Ellis and Tavakol \cite{CET} that the symmetric 2-index square root of the Bell-Robinson superenergy tensor \cite{Bel,BonSen,Sen1,Sen} would act as the effective energy momentum tensor for free gravitational field. Although it is not possible to extract such a square root of a 4-index tensor in general, it was explicitly shown in \cite{BonSen} that for purely Coulomb-like gravitational fields or purely wave like gravitational fields it is always possible. {  It is interesting to note that such a two index tensor describing the energy momentum of gravity also exists in the  Teleparallel equivalent of General Relativity (TEGR). In this case such tensor a is constructed from the torsion tensor, which   identically vanishes in General Relativity \cite{maluf}.}\\

Even if we accept the above described tensor as the energy momentum tensor for free gravity, the next important question is: which tensor can describe the mechanism of energy transfer between free gravity and matter. Clearly this cannot be the energy momentum tensor of matter as that is always conserved. In this paper we explicitly construct such an interaction tensor for free gravity and matter (both in Coulomb-like and wave-like fields), that is purely made up of matter variables and identically vanishes in a  vacuum. However in non-vacuum scenario this gives a new set of conservation laws which states that the sum of energy momentum tensor of free gravity and this interaction tensor is always conserved.\\

 This description of the interplay between free gravity and matter thus generalises the famous \dum and \dee  thought experiment in Newtonian gravity, which we will describe in the next section.  Section 3 discusses generalising that result to General Relativity.  {We then give two examples: } Section 4 discusses how this works in LRS Type II spacetimes, where gravitational induction may be expected to take place, and Section 5 presents the case of Kundt's Petrov Type N spacetimes, which represent gravitational waves. Section 6 summarizes the outcome.

\section{Gravitational Induction thought experiment in Newtonian gravity}
 
This thought experiment, presented by Bondi  and MacCrea \cite{bondi} (see also \cite{jvn} for a beautiful explanation), involves two intelligent and physically identical spherically shaped creatures who are made up of pliable material, that allows them to change their shape. Bondi named these creatures \dum and \dee after the characters of Lewis Carrol's novel {\em Through the looking glass} \cite{Carroll}. They are ordered to move around each other under their mutual gravitational force along highly eccentric orbit in such a way that their centre of gravity remains fixed in space. The rules of the game are as follows:
\begin{itemize}
\item The given orbits are fixed, that is no matter what, they have to move along those orbits.
\item They are allowed to change their shape if required, however, the shapes have to be always symmetrical about the axes perpendicular to the plane of their orbits.
\end{itemize}

Now, since \dum and \dee are pliable, their shape will get distorted by the tidal force of their mutual gravity. Owning to the high eccentricity of their orbits, when they come very close to each other, the strong tidal forces will force their initial spherical shape to become oblate spheroidal. But, if both becomes oblate spheroidal, then the mutual gravitation force between them will increase and that will tend to alter their orbits, which is not allowed by the rules. \\

One way to overcome this problem is as follows. If one becomes oblate the other should become prolate to the same degree and vice-versa, so the mutual force between them remains the same. Now \dum is quite clever but \dee is naive. \dum makes \dee sign a contract stating that \dum will take the initiative of this shape changing operation and \dee will follow suit. In other words, when \dum becomes oblate, \dee will become prolate and when \dum becomes prolate \dee will become oblate, in order to keep their mutual gravitational foci unchanged. \\

Thus, moving in their highly eccentric orbits, when they come close to each other, \dum allows himself to become oblate. To achieve this, he doesn't have to do any work, the tidal force does the work for him. While at this stage, according to the contract, \dee has to become prolate and thus has to do a lot of work against the strong tidal force. On the other hand, when both of them are farthest from each other, \dum changes himself to a prolate shape, he has to do some work against relatively much weaker tidal force, and \dee lets herself become oblate and gains some energy, which is much less than the energy spent before. \\

In a nutshell, therefore after each complete rotation, \dum is gaining a lot of internal energy as the external tidal force is doing work on him, while \dee is losing a lot of internal energy as she is doing work against the external force. This is an excellent example of how the internal energy of a system can be  transferred to another system via {\em gravitational induction}, in Newtonian gravity (there is nothing wavelike about the process).

\section{Is the general relativistic counterpart possible?}

The obvious question one would ask now is, whether a similar experiment is possible within the context of general relativity. Of course, in general relativity, gravity can transfer energy via two mechanisms: Firstly by  gravitational induction, which is similar to the Newtonian case,  governed by the electric part of the Weyl tensor
\be\label{Eab}
E_{ab}=C_{acbd}u^cu^d\;,
\ee
where $u^a$ is the timelike normalised 4-velocity of the fluid; the magnetic part $H_{ab}$ of the Weyl tensor plays no role. This is for example the way the Moon causes tides on Earth. The second mechanism is gravitational radiation (which has no counterpart in Newtonian gravity), that is allowed by the magnetic part of the Weyl tensor
\be\label{Hab}
H_{ab} =\sfr12\ep_{ade}C^{de}{}_{bc}u^c \;,
\ee 
where $ \ep_{abc}=\eta_{abcd}u^d$ is the volume element of the spacelike 3-space perpendicular to $u^a$. In a perturbed Robertson-Walker spacetime, $E_{ab}$ and $H_{ab}$ satisfy wave equations in close analogy to the way $E_a$ and $H_a$ do in the case of electromagnetic theory \cite{hawking66,Ell71,MaaBass}, allowing wave solutions with $E_{ab} \neq 0, H_{ab} \neq 0$. \\

The earliest attempt at investigating gravitational induction in general relativity was by Levy \cite{levy},  who investigated the near-field transfer of gravitational energy for quasi-static axisymmetric
systems by a perturbation method. He defined a relativistic analogue of Bondi's Newtonian Poynting vector and concluded that a quasi-static axisymmetric
system can lose energy only in the presence of a receiver.\\

Just a few years later another famous thought experiment was presented by Penrose \cite{penrose}, which is now 
commonly known as {\it Penrose mechanism}. In this process, energy can be extracted from a rotating black hole. That extraction is made possible because the rotational energy of the black hole is located not inside the event horizon of the black hole, but on the outside of it in a region of the Kerr spacetime called the ergosphere, a region in which a stationary particle  necessarily moves in 
 concurrence with the rotating spacetime. In the Penrose precess, a particle enters into the ergosphere of the black hole, and once it enters the ergosphere, it is forcibly split into two parts, one falls into the black hole and the other escapes from it. With careful arrangement, the escaping particle can be made to have greater mass-energy than the original particle. Although momentum is conserved, more energy can be extracted than was originally provided, the difference being provided by the black hole itself, which loses mass. \\

However both of the above studies were restricted to very specific type of systems, and no general mechanisms were provided that can generalise the concept of energy exchange of free gravity and matter, so as to generalise \dum and \dee to general relativity. Since in general relativity any conservation or exchange of energy or momentum is depicted by the covariant divergence of tensors, we had two important hurdles in this investigation:
\begin{enumerate}
\item What is the energy momentum tensor of a free gravitational field (gravitational field in vacuum)?
\item If we find a energy momentum tensor of a free gravitational field, with which tensor does it interact when energy is transferred between bodies? Clearly this cannot be the energy momentum tensor of matter, which is always conserved according to the Einstein's field equations.
\end{enumerate}

The answer to the first hurdle was provided in steps. First, Maartens and Basset \cite{MaaBass} presented a very interesting result. They proved that the Bell-Robinson tensor \cite{bell1,Bel,bell2,robin}
\be\label{BR}
T_{abcd}\equiv\frac14\left(C_{eabf}C^{e~~~f}_{~cd}+C^*_{eabf}C^{*e~~~f}_{~cd}\right)
\ee
is a unique Maxwellian tensor that can be constructed from the Weyl tensor so as to act as the `super energy-momentum' tensor for gravitational fields (see also \cite{Sen1}), with only one caveat, that the dimension of this tensor is $L^{-4}$ and not $L^{-2}$ like the energy momentum tensor should be.  Based on this result Clifton, Ellis and Tavakol \cite{CET} proposed that the symmetric 2-index square root $ \mathcal{T}_{ab}$, of the Bell-Robinson tensor would act as the effective energy momentum tensor for free gravitational field. Although the square root may not exist for general spacetimes, it was shown that a unique square root exists (up to an arbitrary trace part) for spacetimes that are Petrov type D, which has Coloumb like gravitational field, or N, that is a wave like gravitational field \cite{stephani}. 

In the next two sections, we will try to answer the second question posed above. We will transparently illustrate that  it is possible to construct a matter interaction tensor in both Coloumb-like gravitational fields in locally rotationally symmetric (LRS) class II spacetimes \cite{EllisLRS} (like spherically symmetric spacetimes), which are a subclass of Petrov type D, and in non-expanding and non rotating Kundt class spacetimes \cite{kundt}, (which are a subclass of Petrov type N). This tensor clearly encodes the process of energy and momentum exchange between free gravity and matter. 

\section{Free gravity and matter interaction in LRS-II spacetimes}
We know that for any LRS spacetime \cite{EllisLRS}, apart from the timelike direction $u^a$, there is a preferred spacelike direction $e^a$ at every point of the spacetime manifold, with $u^ae_a=0$. Hence the four dimensional metric can be dimensionally decomposed in a  $1+1+2$ manner along these directions \cite{chris1,clarksonlrs}:
\be
g_{ab}=-u_au_b+h_{ab}=-u_au_b+e_ae_b +N_{ab}\,,
\ee
where $N_{a b}=h_{ab}-e_ae_b=g_{a b}+u_au_b-e_ae_b$ is the projection tensor onto the two dimensional sheet perpendicular to $u^a$ and $e^a$. LRS II spacetimes are a subclass of LRS spacetimes with identically vanishing rotation and spatial twist. Hence in this subclass the magnetic part of the Weyl tensor also vanishes identically ($H_{ab}=0$). To show the interaction in a transparent fashion, let us consider the matter to be of the form of a perfect fluid, that is
\be 
T_{ab}=(\mu+p)u_au_b + p g_{ab}\,,
\ee
where $\mu=T_{ab}u^au^b$ is the energy density and $p=\sfr13T^{ab}h_{ab}$ is the isotropic pressure. We note that in the perfect fluids, the heat flux $Q=e^ah_{ab}T^{bc}u_c=0$, and also the anisotropic stress $\Pi= 0$. The symmetry of LRS-II spacetime allows us to decompose all the kinematical geometrical variables, Weyl variables and matter variables into the following set of scalars:
\be
\{\Theta,\udot,\Sigma, \phi, {\cal E},\mu,p \}\,
\ee
where $\Theta$ is the expansion of $u^a$, $\Sigma\equiv\sigma_{ab}e^ae^b$ is the shear scalar of $u^a$, $\udot$ is the acceleration of $u^a$, and $\phi$ is the expansion of $e^a$. Furthermore ${\cal E}\equiv E_{ab}e^ae^b$ is the electric part of Weyl scalar. The matter variables are: $\mu,p$, which vanish in vacuum spacetimes.
We note that the choice of the timelike vector naturally defines 
two derivatives: the vector $ u^{a}$ is used to define the \textit{covariant
time derivative} along the observers' worldlines (denoted by a dot)
for any tensor $ S^{a..b}{}_{c..d}$, given by 
\be
\dot{S}^{a..b}{}_{c..d}{} = u^{e} \nab_{e} {S}^{a..b}{}_{c..d}\;,
\ee
and the tensor $ h_{ab} $ is used to define the fully orthogonally
\textit{projected covariant derivative} $D$ for any tensor $
S^{a..b}{}_{c..d} $: 
\be D_{e}S^{a..b}{}_{c..d}{} = h^a{}_f
h^p{}_c...h^b{}_g h^q{}_d h^r{}_e \nab_{r} {S}^{f..g}{}_{p..q}\;.
\ee 
In 1+1+2 decomposition, apart from the `{\it time}' (dot) derivative, of a tensor, we naturally get two new derivatives, 
which $ e^{a} $ defines. For any tensor $ S_{a...b}{}^{c...d}  $: 
\ba
\hat{S}_{a..b}{}^{c..d} &\equiv & e^{f}D_{f}S_{a..b}{}^{c..d}~, 
\\
\delta_fS_{a..b}{}^{c..d} &\equiv & N_{a}{}^{f}...N_{b}{}^gN_{h}{}^{c}..
N_{i}{}^{d}N_f{}^jD_jS_{f..g}{}^{i..j}\;.
\ea 
The hat-derivative is the derivative along the $e^a$ vector-field in the surfaces orthogonal to $ u^{a} $. The $\delta$ -derivative is the projected derivative onto the sheet, with the projection on every free index. The symmetry of LRS spacetimes makes the $\delta$ -derivative of any scalar vanish.
For the LRS-II spacetimes the full covariant derivatives of the preferred timelike and spacelike vectors are:
\be\label{dele}
 \del_b\e^a=-\udot u_bu^a+\bra{\Sigma+\sfr13\Theta}\e_b u^a+\frac{1}{2}\phi\N^a_b\;,
\ee
\be\label{delu}
\del_bu^a=-\udot u_be^a+\bra{\sfr13\Theta+\Sigma}\e_b\e^a+\bra{\sfr13\Theta-\sfr12\Sigma}\N^a_b\;.
\ee

The field equations in terms of these scalar variables are given as\\

\textit{Propagation}:
\ba
\hat\phi  &=&-\sfr12\phi^2+\bra{\sfr13\Theta+\Sigma}\bra{\sfr23\Theta-\Sigma}\nonumber\\&&-\sfr23\mu
    -\E,\label{hatphinl}
\\  
\hat\Sigma-\sfr23\hat\Theta&=&-\sfr32\phi\Sigma\,,\label{Sigthetahat}
 \\  
\hat\E-\sfr13\hat\mu&=&
    -\sfr32\phi\E\;.
    \label{ehat}
\ea

\textit{Evolution}:
\ba
   \dot\phi &=& -\bra{\Sigma-\sfr23\Theta}\bra{\udot-\sfr12\phi}\ , \label{phidot}
\\   
\dot\Sigma-\sfr23\dot\Theta
&=&
-\udot\phi+2\bra{\sfr13\Theta-\sfr12\Sigma}^2\nonumber\\
        &&+\sfr13\bra{\mu+3p}-\E\, ,\label{Sigthetadot}
\\  
\dot\E -\sfr13\dot \mu &=&
    \bra{\sfr32\Sigma-\Theta}\E
     -\sfr12\bra{\mu+p}\bra{\Sigma-\sfr23\Theta}. \label{edot}
\ea

\textit{Propagation/evolution}:
\ba
   \hat\udot-\dot\Theta&=&-\bra{\udot+\phi}\udot+\sfr13\Theta^2
    +\sfr32\Sigma^2 \nonumber\\
    &&+\sfr12\bra{\mu+3p}\ ,\label{Raychaudhuri}
\\
    \dot\mu&=&-\Theta\bra{\mu+p},\,  \label{Qhat}
\\    
\hat p&=&-\bra{\mu+p}\udot\ . \label{Qdot}
\ea
\\

 The above set of equations are  equivalent to the Einstein field equations, but are covariantly written in terms of the geometrical and matter variables instead of the metric functions. The matter conservation equations are implied by the Einstein equations via the doubly contracted Bianchi Identities. We can easily see that equations (\ref{Qhat}) and (\ref{Qdot}) are the matter conservation equations.

The 3-Ricci scalar of the spacelike 3-space orthogonal to $u^a$ can be expressed as
\be
^3R =-2\bras{\hat \phi +\sfr34\phi^2 -K}\;,\label{3sca}
\ee
where $K$ is the Gaussian curvature of the 2-sheet defined by
$^2R_{ab}=KN_{ab}$. In terms of the covariant scalars we can write the Gaussian curvature $K$ as
\be
K = \sfr13\mu-\E+\sfr14\phi^2
-\bra{\sfr13\Theta-\sfr12\Sigma}^2\ . \label{gauss}
\ee
Finally the evolution and propagation equations for the Gaussian curvature $K$ are
\ba
\dot K &=& -\bra{\sfr23\Theta-\Sigma}K\ , \label{Kdot}\\
\hat K &=& -\phi K\ . \label{Khat}
\ea
\subsection{Energy momentum tensor of free gravitational field}

As described in \cite{CET,boos}, for Coulomb-like gravitational fields that are  described by a Petrov type D spacetime (of which LRS-II is a subclass), the symmetric 2 index square root of the Bell-Robinson tensor $\Tf^G_{ab}$ can be written as 
\begin{equation}
\label{TtypeD}
\Tf^G_{ab} = \al\left[ 3 \vert \Psi_2 \vert \left( m_{(a} \bar{m}_{b)} + l_{(a} k_{b)} \right) + f g_{ab} \right],
\end{equation}
where $\al$ is a constant, $l_a$ and $n_a$ are real null vectors, and $m_a$ and $\bar{m}_a$ are complex null vectors at a given point in the spacetime in the Newman-Penrose tetrad formalism. $\Psi_2$ is the only non zero complex Weyl scalar for Petrov type D spacetimes and  `$f$' is a free function that can be fixed via  physical requirements. 
In terms of electric and magnetic parts of Weyl tensor the complex Weyl scalar can be written as 
\be\label{psi2}
\vert\Psi_2\vert=\sqrt{\frac{2W}{3}}\,,
\ee
where 
\be\label{W}
W=\frac14\LB E_{ab}E^{ab} + H_{ab}H^{ab}\RB.
\ee
In terms of the semitetrad decomposition of LRS-II spacetimes, the Newman Penrose null tetrads are given as 
\be
k^a=\frac{1}{\sqrt{2}}\bra{u^a+e^a},\; l^a=\frac{1}{\sqrt{2}}\bra{u^a-e^a},\;N_{ab}=2m_{(a}\bar{m}_{b)}\;.
\ee
The square of the electric part of the Weyl tensor becomes
\be
E_{ab}E^{ab}=\frac32{\cal E}^2\;.
\ee
Since $H_{ab}=0$, we get
\be
W=\frac38{\cal E}^2\,,
\ee
and hence the non-zero Weyl scalar in NP formalism can be written in terms of 1+1+2 scalar as
\be
\vert\Psi_2\vert =\pm\frac12{\cal E}\;.
\ee
Therefore, the effective energy momentum tensor for free gravitational field in LRS-II spacetimes can be written as 
\begin{equation}
\label{TypeD1+1}
 \Tf^G_{ab}= \al\left[\frac32{\cal E}\left(u_au_b-e_ae_b\right)+\LB f+\frac34{\cal E}\RB g_{ab}\right].
\end{equation}
Then the divergence of the the above tensor can be easily calculated using equations (\ref{dele}) and (\ref{delu}), which is given by,
\ba\label{divtab}
\nabla^b\Tf^G_{ab}&=&\frac32\al\LB\dot{\E}u_a +\E\udot e_a+\E\Theta u_a\RB\nonumber\\
&&-\frac32\al\LB\hat{\E}e_a+\E\LB\Sigma+\sfr13\Theta\RB u_a+\E\LB\udot+\phi\RB e_a\RB\nonumber\\
&& +\al g_{ab}\nabla^b f + \frac34\al g_{ab}\nabla^b \E\;.
\ea
The components of the above vector along $u^a$ and $e^a$ are then 
\be
u^a\nabla^b\Tf^G_{ab}=-\frac34\al\dot{\E}+\frac32\al\E\LB\Sigma-\sfr23\Theta\RB+\al\dot{f}\;,
\ee
\be
e^ah_a^c\nabla^b\Tf^G_{ac}=-\frac34\al\hat{\E}+\frac32\al\E\phi+\al\hat{f}\;.
\ee
If we assume the physically reasonable condition that this effective energy momentum tensor $ \Tf^G_{ab}$ is conserved (divergence free) in vacuum, then comparing the above two equations with field equations (\ref{edot}) and (\ref{ehat}) in vacuum, we see that the function `$f$' must satisfy the following equations
\be
\dot{f}=-\frac14\dot{\E}\;\;,\;\; \hat{f}=-\frac14\hat{\E}\;.
\ee
The above equations uniquely determine the function $f$ as 
\be
f=-\frac14\E +\lambda_1,
\ee
where $\lambda_1$ is arbitrary constant. Hence the final energy momentum tensor for the free gravity in LRS II spacetimes becomes
\be\label{TfLRS}
\Tf^G_{ab}=\al\left[\frac32\E\left(u_au_b-e_ae_b\right)+\LB\frac12\E+\lambda_1\RB g_{ab}\right]\;,
\ee
with
\be
\mu^G=\al(\E+\lambda_1),\;p^G=\al\lambda_1,\;\Pi^G=-\al\E,\;Q^G=0\,.
\ee

\subsection{Interaction of free gravity with matter}

Let us now consider the case when the free gravity interacts with matter. Then from equations (\ref{ehat}) and (\ref{edot}), we can immediately see that $ \Tf^G_{ab}$ is no longer conserved, that is, $\nabla^b\Tf^G_{ab}\ne 0$. This is not surprising as free gravity must exchange energy and momentum with matter. However here we come to a paradoxical situation as energy momentum tensor of standard matter is always conserved, that is, $\nabla_bT^{ab}=0$. This brings us to this important question:\\

{\it With which tensor does free gravity interact?}\\

In other words do equations (\ref{edot}) and (\ref{ehat}) represent any conservation law? For arguments sake, let's suppose they do and they can be written in the component  form along 
 $u^a$ and $e^a$ in the following way:
 \be\label{div1}
 u^a\nabla^b(\Tf^G_{ab}+\Td^I_{ab})=0=e^a\nabla^b(\Tf^G_{ab}+\Td^I_{ab})\;,
 \ee
where $\Td^I_{ab}$ is a general symmetric tensor of rank 2, representing the free gravity and matter interaction, via gravitational induction, and is given as
\be\label{Tdgen}
\Td^I_{ab}=Lu^au^b+\mathcal{Q}(u_ae_b+e_au_b)+M e_ae_b +Ng_{ab}\;.
\ee
Now calculating the divergence (\ref{divtab}) and comparing that with equations (\ref{edot}) and (\ref{ehat}), we get 
\ba\label{Td}
\Td^I_{ab}&=&\al\left[-\frac12(\mu+p)\LB u_au_b-e_ae_b\RB+\mathcal{Q}(u_ae_b+e_au_b)\right.\nonumber\\
&&\left.-\frac16(\mu+3p)g_{ab}\right]\;,
\ea
with the function $\mathcal{Q}$ satisfying the following differential equations:
\be
\dot{\mathcal{Q}}+\left(\Sigma+\sfr43\Theta\right)\mathcal{Q}+\frac12(\mu+p)\phi=0\;,
\ee
and
\be
\hat{\mathcal{Q}}+2\mathcal{Q}\LB\udot+\sfr12\phi\RB=0\,.
\ee
This clearly shows, that there is indeed a free-gravity matter interaction tensor $\Td^I_{ab}$, whose components are completely determined by the matter variables. In fact the thermodynamic variables associated with this interaction tensor can be written as
 \be
\mu^I=-\frac13\al\mu\;,\;p^I=-\frac16\al p\;, 
\ee
\be
\Pi^I=\frac16\al(2\mu+p)\;,\;Q^I=\mathcal{Q}\,.
\ee
This tensor interacts with free gravity in such a way that 
\be
\nabla^b(\Tf^G_{ab}+\Td^I_{ab})=0\;,
\ee
that is, the sum of the free gravity energy momentum tensor and the interaction tensor is always conserved in LRS-II spacetimes. Thus for these spacetimes $\Td^I_{ab}$ and $\Tf^G_{ab}$ acts as the tensor enabling \dum and \dee to exchange energy and momentum via gravitational induction. Let us now discuss a few limiting cases:
\begin{enumerate}
\item It is obvious that in vacuum the interaction tensor vanish identically, with vanishing $\mu^I$, $p^I$ and $\Pi^I$. In this case equations for $\mathcal{Q}$ becomes homogeneous differential equations for which $\mathcal{Q}=0$ is obviously a solution.
\item For conformally flat spacetimes (like FLRW) when the Weyl tensor vanish identically, we can easily see that the energy momentum tensor for the free gravity vanish ($\Tf^G_{ab}=0$). In this case the interaction tensor has nothing to interact with and is conserved ($\nabla^b\Td^I_{ab}=0$). However we can easily check that these conservation equations   give the same information as the conservation of standard matter ($\nabla^bT_{ab}=0$). Hence in this case the interaction tensor becomes redundant.
\end{enumerate}

\section{Free gravity and matter interaction in non expanding and non rotating gravitational wave spacetimes}

The previous section generalised the concept of \dum and \dee to the spacetimes that allows gravitational induction but no gravitational waves, due to the vanishing of the magnetic part of the Weyl tensor. To see the \dum and \dee effect in gravitational waves, let us now consider the other extreme: the case of plane-fronted transverse gravitational waves (that have been detected by LIGO). The geometry associated with these waves is close to that of Kundt's class \cite{kundt,stephani}, which are the class of all Petrov type N solutions with vanishing expansion and rotation (the waves detected by LIGO will have a very small but nevertheless non-zero expansion). These types of waves are closely analogous to our understanding of electromagnetic waves, and fit in with the idea of gravitational wave fronts, as the congruence of null curves they follow is irrotational and hence hyper-surface orthogonal, and so may be used as good exact gravitational wave representatives \cite{PodOrt}. \\

To get a transparent geometrical feeling for this class of spacetimes, the semitetrad $1+1+2$ covariant formalism, described in the previous section, is not very suitable. Hence in this section we use the Newman-Penrose formalism \cite{NP}. We assume that the spacetime is spanned by the Newman Penrose (NP) null tetrad $(l^a,n^a,m^a,\mb^a)$. Here  $l_a$ and $n_a$ are real null vectors, $m_a$ and $\bar{m}_a$ are the complex null vectors at a given point in the spacetime and they satisfy the following properties: $l_al^a=n_an^a=m_am^a=\bar{m}_a\bar{m}^a=0$, $l_an^a=-1,m_a\bar{m}^a=1$ and $l_am^a=n_am^a=l_a\bar{m}^a=n_a\bar{m}^a=0$.
The metric in terms of these tetrads is given by
\be
g_{ab}=-2l_{(a}n_{b)}+2m_{(a}\mb_{b)}\,.
\ee
We can define the directional derivative along each of the tetrad vector fields in the following way:
\be
D\equiv l^a\nabla_a,\; \Delta\equiv n^a\nabla_a,\; \delta\equiv m^a\nabla_a,\; \overline{\delta}\equiv \overline{m}^a\nabla_a\,.
\ee
Also the following are the NP spin coefficients:
\be
\left\{\kappa,\sigma,\nu,\lambda,\tau,\rho,\pi,\mu_{np},\epsilon,\gamma,\alpha,\beta\right\}\;.
\ee
In terms of these spin coefficients the full covariant derivatives of the null vectors are given by
\ba\label {delbla}
\nabla_bl^a&=& -(\gamma+\ov{\gamma})l_bl^a+\ov\tau l_bm^a + \tau l_b\mb^a \nonumber\\
&&-(\ep+\ov\ep)n_bl^a + \ov\kappa n_bm^a +\kappa n_b\mb^a \nonumber\\
&&+(\alpha+\ov\beta)m_bl^a-\ov\sigma m_bm^a -\rho m_b\mb^a \nonumber\\
&&+(\ov\alpha+\beta)\mb_bn^a-{\ov{\rho}}\, \mb_bm^a-\sigma\mb_b\mb^a\,.
\ea
\ba\label {delbna}
\nabla_bn^a&=& (\gamma+\ov{\gamma})l_bn^a-\nu l_bm^a -\ov\nu l_b\mb^a \nonumber\\
&&+(\ep+\ov\ep)n_bn^a -\pi n_bm^a -\pi n_b\mb^a \nonumber\\
&&-(\alpha+\ov\beta)m_bn^a+\lambda m_bm^a +\ov\mu_{np} m_b\mb^a \nonumber\\
&&-(\ov\alpha+\beta)\mb_bn^a+\mu_{np}\mb_bm^a-\ov\lambda\,\mb_b\mb^a\,.
\ea
\ba\label {delbma}
\nabla_bm^a&=& \ov\nu l_bl^a+\tau l_bn^a -(\gamma-\ov\gamma) l_bm^a \nonumber\\
&&+\ov\pi n_bl^a -\kappa n_bn^a -(\ep-\ov\ep) n_bm^a \nonumber\\
&&-\ov{\mu}_{np}m_bl^a-\rho m_bn^a +(\alpha-\ov\beta) m_bm^a \nonumber\\
&&-\ov\lambda\mb_bl^a-\sigma\mb_bn^a-(\ov\alpha-\beta),\mb_bm^a\,.
\ea
Once again we consider the perfect fluid form of matter
\be
T_{ab}=(\mu+p)u_au_b + p g_{ab}\,,
\ee
which directly gives the Ricci tensor  $R_{ab}=T_{ab}-\sfr12g_{ab}T$
\be
R_{ab}=(\mu+p)u_au_b + \frac12(\mu-p) g_{ab}\,.
\ee
In terms of the NP tetrads, the above can be written as
\ba\label{Ricci}
R_{ab}&=&\frac12(\mu+p)(l_al_b+n_an_b)+p(l_an_b+n_al_b)\nonumber\\
&&+\frac12(\mu-p)(m_a\mb_b+\mb_am_b)\;.
\ea
Then the non-zero tetrad components of the Ricci tensor are:
\be
\Phi_{00}=\frac12R_{ab}l^al^b=\frac14(\mu+p)\,,
\ee
\be
\Phi_{11}=\frac14R_{ab}(l^an^b+l^bn^a)=\frac18(\mu+p)\,,
\ee
\be
\Phi_{22}=\frac12R_{ab}n^an^b=\frac14(\mu+p)\,,
\ee
\be
\Lambda= \frac{R}{24} =\frac{1}{24}(\mu-3p)\;.
\ee
Now , for the Kundt class of spacetime considered in this section we, the symmetry makes the following spin coefficients vanish
\be\label{spin0}
\sigma=\kappa=\pi=\epsilon=\rho=0\;.
\ee
Also the only non-zero Weyl scalar is $\Psi_4$. We will require the following Bianchi identity for $\Psi_4$:
\be\label{bian1}
D\Psi_4=\lambda\Phi_{22}\,.
\ee
\subsection{Energy momentum tensor for free gravitational field}

Again, as described in \cite{CET}, for wave-like gravitational fields that are described by Petrov type N spacetimes (of which the Kundt class is a subclass),  the symmetric 2 index square root of the Bell-Robinson tensor can be written as 
\be
\label{TtypeN}
\Tf^G_{ab} = \bl \vert \Psi_4 \vert l_al_b + f g_{ab}\,
\ee
where $\bl$ is a constant and the function `$f$' is a free function, that can be fixed via certain physical requirements. Calculating the divergence of the above tensor we get
\ba
\nabla^b\Tf^G_{ab}&=&\bl l_a D\Psi_4+ \bl\Psi_4 Dl_a 
+\bl\Psi_4l_a\nabla^bl_b+g_{ab}\nabla^bf\;.
\ea
Demanding that the above should vanish in vacuum and using equation (\ref{delbla}), (\ref{spin0}) and (\ref{bian1}), we get
\be
\nabla^bf=0\;.
\ee
That is the trace $f$ is just a constant which can be taken to be zero without any loss of generality. Therefore the energy momentum tensor for free gravity in this class of spacetimes become
\be
\label{TtypeN}
\Tf^G_{ab} = \bl \vert \Psi_4 \vert l_al_b \,.
\ee

\subsection{Interaction of free gravity and matter}

Let us now suppose that we inject matter into the vacuum spacetime. The free gravity will interact with matter and the energy momentum tensor of free gravity will no longer be conserved. In fact in the presence of matter the divergence of $\Tf^G_{ab}$ is given by
\be\label{div2}
\nabla^b\Tf^G_{ab}=\bl\frac{\lambda}{4}(\mu+p)l_a\;.
\ee
Just by inspection we can guess the form of free gravity matter interaction tensor $\Td^G_{ab}$ as
\be
\Td^G_{ab}=\bl \mathcal{Q} l_al_b,
\ee
where $\mathcal{Q}$ is the solution of the equation
\be
D\mathcal{Q}=-\frac{\lambda}{4}(\mu+p)\;.
\ee
We can immediately see that 
\be
\nabla^b(\Tf^G_{ab}+\Td^I_{ab})=0\;,
\ee
that is, the sum of the free gravity energy momentum tensor and the interaction tensor is conserved even in the case of a pure radiation like field, thus the \dum and \dee effect is present here too. Also, even in this case the interaction tensor vanishes in vacuum, as $\mathcal{Q}=0$ is a solution to the homogeneous equation $D\mathcal{Q}=0$.

\section{The story so far}

Let us now summarise what we have shown so far. We know the Riemann curvature tensor $R^a_{~~bcd}$ contains all the information about the geometry of the spacetime manifold. This tensor can be decomposed in the Ricci part and the traceless Weyl part. The former can be equated to the matter via Einstein field equations, while the later defines the free gravitational field. 
Therefore any interaction between free gravity and matter must be encoded in the Bianchi identities
\be\label{bian2}
\nabla_eR^a_{~~bcd}+\nabla_dR^a_{~~bec}+\nabla_cR^a_{~~bde}=0\;.
\ee
Now in terms of the double dual of Riemann tensor
\be
{}^*R_{abcd}^*=\frac14\eta_{absm}\eta_{cdpq}R^{smpq}\;,
\ee
the above equation (\ref{bian2}) can be written as a single vanishing divergence equation \cite{MTW}
\be\label{bian3}
\nabla^a{}^*R_{abcd}^*=0\;.
\ee
Now it turns out that, when the spacetime has certain symmetries, for example purely Coulomb like Petrov type D spacetimes or purely radiation like Petrov type N spacetimes, this single vanishing divergence of a tensor of type (0,4) can be decomposed into two sets of vanishing divergence equations of tensors of type (0,2). The first set
\be
\nabla^bG_{ab}=0=\nabla^bT_{ab}\;,
\ee
is always present irrespective of any symmetries and denotes the usual conservation of standard matter, while the second set 
\be
\nabla^b(\Tf^G_{ab}+\Td^I_{ab})=0\;,
\ee
is 
 present at least in Type D and N spacetimes and denotes the interplay between free gravity and matter. Given the two index square root of Bel Robinson tensor as the energy momentum tensor for free gravity, the components of the interaction tensor is determined by matter thermodynamic variables, as well as local spacetime integrals of these entities. Thus the interaction tensor is related to more usual quantities not algebraically or differentially, but through an integral.
In type D solutions, $\Psi_2 = \sqrt{ 2/3(E^{ab}E_{ab}+H^{ab}H_{ab})} \ne 0$, 
and in type N solutions, $\Psi_4= \sqrt{4(E^{ab}E_{ab}+H^{ab}H_{ab})} \ne 0$. 
We considered LRSII (type D) where $H_{ab}$ is equal to zero
and Kundt class (type N) where both $E_{ab} \neq 0$ and $H_{ab} \neq 0.$
So the first case is entirely "induction" and second case entirely "radiation". This generalises \dum and \dee to both general relativistic scenarios. Lastly, we would also like to emphasise here that that our proposal for the interaction tensor crucially hinges on the hypothesis that the energy momentum tensor of free gravity is given by the two index square root of the Bell Robinson tensor \cite{CET}. Given this hypothesis, we clearly demonstrated that is it possible to extract a new set of conservation laws from the Bianchi Identities, and propose that such conservations laws are indeed physically meaningful (as is to be expected of any conservation laws). This new conservation law is separate from usual energy-momentum conservation, and may only apply to situations with extra symmetry.\\

\par
\noindent{\bf Acknowledgments}\\
RG and GE thank the South African National Research Foundation (NRF) for support, and GE thanks the University of Cape Town Research Committee (URC) for financial  support. We thank two referees for helpful comments.

\section*{References} 

\vspace{0.1in}

\end{document}